\documentclass[doublecol]{epl2}
\usepackage{amssymb}
\usepackage{bbm}
\usepackage{indentfirst}
\usepackage{graphicx}
\usepackage{amsmath,amssymb}
\usepackage{bpchem}
 
\usepackage{color}
\usepackage{tabularx}
\usepackage{multirow}
\usepackage{ulem}
\usepackage{makecell}
\usepackage{array}
\usepackage{appendix}

\usepackage{multirow}
\usepackage{float} 
\usepackage[colorlinks=true,citecolor=blue,linkcolor=blue,urlcolor=blue]{hyperref}

\title{Dark-state solution and symmetries of the two-qubit multimode
asymmetric quantum Rabi model}
\shorttitle{Dark-state solution and symmetries of the two-qubit multimode
asymmetric quantum Rabi model} 
\author{Ze-Feng Lei\inst{1} \and Junlong Tian\inst{2} \and Jie Peng\inst{1}\thanks{E-mail: \email{jpeng@xtu.edu.cn}}}
\shortauthor{Ze-Feng Lei \etal}
\institute{                    
  \inst{1} Hunan Key Laboratory for Micro-Nano Energy Materials and Devices and School of Physics and Optoelectronics, Xiangtan University, Hunan 411105, China\\
  \inst{2} Department of electronic science, College of Big Data and Information Engineering, Guizhou University, Guiyang 550025, China
}
\pacs{}{}

\abstract{
We study the two-qubit asymmetric quantum Rabi model (AQRM) and find its dark-state solution. Such solutions have at most one photon and constant eigenenergy in the whole coupling regime, causing level crossings in the spectrum, although there is no explicit conserved quantity except energy. We find an operator in the eigenenergy basis to label all the degeneracies with its eigenvalues, and compare it with the well-known hidden symmetry which exists when bias parameter $\epsilon$ is a multiple of half of the resonator frequency $\omega$. Extended to the multimode case, we find symmetries related with conserved bosonic number operators, which also cause level crossings. This provides a perspective for symmetry studies on generalized Rabi models.}

\begin{document}

\maketitle

\section{Introduction}\label{sec:01}
The quantum Rabi model (QRM) \cite{ipt} describes the interaction between a single-mode cavity and a qubit. It has wide applications in quantum optics \cite{pte,rmb,dsc,sbp,qrm,daq}, circuit quantum electrodynamics (QED)  \cite{zpi,nem,dau,suc,cqe}, cavity QED \cite{sca,eom,qdc,ccr,aan,sic}, quantum information \cite{dom,qga,tpe} and so on \cite{sep,RG,RJP}. Its semiclassical form was first introduced by Rabi \cite{otp,sqi}. Jaynes and Cummings \cite{ipt} carried out the rotating wave approximation (RWA) \cite{sdo} for the QRM, and obtained the analytical solution in 1963. However, the ultrastrong \cite{NTD} and even deep strong coupling \cite{YFF} has been realized in experiments, where the RWA fails. The analytical solution to the QRM was found by Braak \cite{iot} in 2011 in the Bargmann \cite{oah} space, and then retrieved by Chen \textit{et al.} with Bogoliubov operator approach  \cite{eso}. There are many interesting studies on QRM and its generalizations \cite{larson,scv,pude1,liu,ying1,ars,qso,cong,pude2,Felicetti,yu1,yu2,duan,ying2,huang,
braak1,li1,li2,li3,tang} recently. Since there is no closed subspace in the photon number space, the eigenstate normally consists of infinite photons, making the dynamics in the ultrastrong coupling regime quite complex. However, there are special dark states with finite photons for the multiqubit and multimode case \cite{SOT,aso,dls,ops}, since the coherent superposition of basis with $N$-photon will cancel the population of higher photon number states when applied by the Hamiltonian. Such solutions exist in the whole qubit-photon coupling regime with constant eigenenergy when $N=1$. Taking advantage of such special dark states, one can fast generate $W$-states \cite{ops}, Bell states \cite{xin}, and high-quality single photon sources \cite{dsp} in the ultrastrong coupling regime deterministically through adiabatic evolution.

Meanwhile, the AQRM has attracted much interest recently. It has an additional static bias term $\epsilon\sigma_x$, which was considered physically as a spontaneous transition of the qubit \cite{iot}. Moreover, the AQRM widely appears in circuit QED systems, where the static bias of the superconducting flux qubit can be tuned externally \cite{oot,uat,ioq,tpq}. This provides more options for precise quantum control of the system. For the AQRM Hamiltonian, the presence of the static bias breaks the $\mathbbm{Z}_2$ symmetry $R=\exp{(i\pi a^\dag a)}\sigma_z$ of the QRM. Hence, generally there is no level crossing in the spectrum. However, recent studies \cite{iot} have found that level crossings are restored when $ \epsilon $ takes half-integer value of $\omega$, indicating a hidden symmetry in the AQRM \cite{hsa,atf,ths}. In addition, many works \cite{ths,GSO} have rigorously constructed the hidden symmetry operators using different methods, and inspired people to study the hidden symmetry of generalized AQRMs \cite{shi,SOO,ASA,HSO}.

Recent hidden symmetry studies of the AQRM focus on the case of $\epsilon$ is a multiple of $\omega/2$ \cite{atf,ths,shi,GSO,wsw}. However, it is interesting to explorer whether there are other kinds of symmetries.
In this paper, we first study the two-qubit AQRM and find a special dark state with at most one photon and constant eigenenergy in the whole coupling regime, corresponding to a horizontal line $E=\omega$ in the spectrum. Apparently, this one-photon solution will bring in level crossings. Similarly,
the level crossings existing when $\epsilon=n\omega/2$ \cite{ths} is also brought by solutions involving only finite basis in the extended coherent state space \cite{eso}, located at the baseline $E=n\omega-g^2/\omega\pm\epsilon$. 
However, the level crossings here only happen between the one-photon solution and other energy levels, so we can use a projection operator in the eigenenergy basis to label all the degeneracy points. Such degeneracy also happens in the parity subspace when $\epsilon=0$. Whether such level crossings due to a hidden symmetry or just fine tuning of parameters is an interesting question to study.  We extend the two-qubit AQRM to the $M$-mode case and introduce a Bogoliubov transformation  \cite{stt} to rewrite the Hamiltonian, so that the dark state solution with at most one photon and constant energy is directly obtained. The hidden symmetry still exists when $ \epsilon $ takes half-integer value of $\omega$ and its symmetry operator is obtained. Moreover, there are other $M-1$ symmetries related with conserved bosonic number operator $b_j^\dag b_j$ for $j=2,\ldots,M$.

\section{Special dark state solution of the two-qubit AQRM and level crossings it caused}\label{s2}
The Hamiltonian of the two-qubit AQRM reads ($\hbar$=1)
\begin{eqnarray}\label{h1}
H&=&\omega a^\dagger a+g_{1}\sigma_{1x}(a^\dagger+a)+g_{2}\sigma_{2x}(a^\dagger+a)+\Delta_1\sigma_{1z}\nonumber\\
&&+\Delta_{2}\sigma_{2z}+\epsilon_{1}\sigma_{1x}+\epsilon_{2}\sigma_{2x},
\end{eqnarray} 
where $ a^\dagger $ and $ a $ are creation and annihilation operators with cavity frequency $ \omega $, respectively. The two qubits are described by Pauli matrices 
$\sigma_x$ and $\sigma_z$ with the energy level  splitting $2\Delta$. $ g_{1} $ and
 $ g_{2} $ are the qubit–photon coupling constants for the two qubits, respectively. 
 $ \epsilon_{1} $ and $ \epsilon_{2} $ are the static bias of the two qubits, respectively.

For this Hamiltonian, the presence of the static bias breaks the $\mathbbm{Z}_2$ symmetry $R=\exp(i\pi a^\dag a)\sigma_{1z}\sigma_{2z}$. The eigenstates generally consist of infinite photon number states. To find possible level crossings beside when $ \epsilon $ takes half-integer value of $\omega$, we search for solution involving finite basis in the photon number space. Meanwhile, finding such solutions will be interesting and useful \cite{ops,dsp} for fast quantum information protocols using ultrastrong coupling. Supposing there is an eigenstate with at most one photon $|\psi\rangle=c_{0,1}|0,g,g\rangle+c_{0,2}|0,e,e\rangle+c_{0,3}|0,e,g\rangle+c_{0,4}|0,g,e\rangle+c_{1,1}|1,g,g\rangle+c_{1,2}|1,e,e\rangle+c_{1,3}|1,e,g\rangle+c_{1,4}|1,g,e\rangle$, then the eigenenergy equation reads ( $ \omega$ is set to 1)
 
\begin{widetext}\fontsize{9pt}{14pt}\selectfont 
\setlength{\arraycolsep}{-10pt}
\begin{equation}\label{E1}
\begin{pmatrix}
   -\Delta_1-\Delta_2-E&0&\epsilon_1&\epsilon_2&0&0&g_{1}&g_{2}\\
    0&\Delta_1+\Delta_2-E&\epsilon_2&\epsilon_1&0&0&g_{2}&g_{1}\\
    \epsilon_1&\epsilon_2&\Delta_1-\Delta_2-E&0&g_{1}&g_{2}&0&0\\
    \epsilon_2&\epsilon_1&0&-\Delta_1+\Delta_2-E&g_{2}&g_{1}&0&0\\
    0&0&g_{1}&g_{2}&1-\Delta_1-\Delta_2-E&0&\epsilon_1&\epsilon_2\\
    0&0&g_{2}&g_{1}&0&1+\Delta_1+\Delta_2-E&\epsilon_2&\epsilon_1\\
    g_{1}&g_{2}&0&0&\epsilon_1&\epsilon_2&1+\Delta_1-\Delta_2-E&0\\
    g_{2}&g_{1}&0&0&\epsilon_2&\epsilon_1&0&1-\Delta_1+\Delta_2-E\\
    0&0&0&0&0&0&\sqrt{2}g_{1}&\sqrt{2}g_{2}\\
    0&0&0&0&0&0&\sqrt{2}g_{2}&\sqrt{2}g_{1}\\
    0&0&0&0&\sqrt{2}g_{1}&\sqrt{2}g_{2}&0&0\\
    0&0&0&0&\sqrt{2}g_{2}&\sqrt{2}g_{1}&0&0\\
  \end{pmatrix} 
\begin{pmatrix}
c_{0,1}\\
c_{0,2}\\
c_{0,3}\\
c_{0,4}\\
c_{1,1}\\
c_{1,2}\\
c_{1,3}\\
c_{1,4}\\
\end{pmatrix}=0,
\end{equation}
\end{widetext}
\begin{floatequation}
\mbox{\textit{see eq.~\eqref{E1} below}}
\end{floatequation}

which requires
\begin{equation}\label{E2}
\begin{normalsize} 
\left|
  \begin{array}{cccc}
  0&0&\sqrt{2}g_{1}&\sqrt{2}g_{2}\\
  0&0&\sqrt{2}g_{2}&\sqrt{2}g_{1}\\
  \sqrt{2}g_{1}&\sqrt{2}g_{2}&0&0\\
  \sqrt{2}g_{2}&\sqrt{2}g_{1}&0&0\\
  \end{array}
\right|=0\\
\end{normalsize} .
\end{equation}
Such that $ g_{2}=\pm g_{1}=g $, $c_{1,1}=\mp c_{1,2}$, $c_{1,3}=\mp c_{1,4}$, and eq. (\ref{E1}) reduces to 

\begin{widetext}\fontsize{10pt}{14pt}\selectfont 
\setlength{\arraycolsep}{-6pt}
\begin{equation}\label{Ee}
\begin{pmatrix}
   -\Delta_1-\Delta_2-E&0&\epsilon_1&\epsilon_2&0&0\\
    0&\Delta_1+\Delta_2-E&\epsilon_2&\epsilon_1&0&0\\
    \epsilon_1&\epsilon_2&\Delta_1-\Delta_2-E&0&0&0\\
    \epsilon_2&\epsilon_1&0&-\Delta_1+\Delta_2-E&0&0\\
    0&0&\pm g& g&1-\Delta_1-\Delta_2-E&\epsilon_1\mp\epsilon_2\\
    0&0& g& \pm g&\mp(1+\Delta_1+\Delta_2-E)&\epsilon_2\mp\epsilon_1\\
    \pm g& g&0&0&\epsilon_1\mp\epsilon_2&1+\Delta_1-\Delta_2-E\\
    g& \pm g&0&0&\epsilon_2\mp\epsilon_1&\mp(1-\Delta_1+\Delta_2-E)\\
  \end{pmatrix} 
\begin{pmatrix}
c_{0,1}\\
c_{0,2}\\
c_{0,3}\\
c_{0,4}\\
c_{1,1}\\
c_{1,3}\\
\end{pmatrix}=0.
\end{equation}
\end{widetext}

\begin{floatequation}
\mbox{\textit{see eq.~\eqref{Ee} below}}
\end{floatequation}

If there are less nonzero rows than columns in the  above $ 8\times6 $ coefficient matrix after elementary row transformation, then there are nontrivial solutions. This can be done when $E=1$, $g_{2}=\pm g_{1}=g$, $\epsilon_2=\pm \epsilon_1=\epsilon$, $\epsilon^2=\dfrac{\Delta_1^{4}+(-1+\Delta_2^{2})^{2}-2\Delta_1^{2}(1+\Delta_2^{2})}{4}\geq0 $, and the coefficient matrix becomes

\begin{eqnarray}
\begin{small} 
\begin{pmatrix}
   1&0&0&0&0&\dfrac{\mp(\Delta_1-\Delta_2)(-1+\Delta_1+\Delta_2)}{2g}\\
    0&1&0&0&0&\dfrac{(\Delta_1-\Delta_2)(1+\Delta_1+\Delta_2)}{2g}\\
    0&0&1&0&0&\dfrac{-\epsilon(\Delta_1-\Delta_2)}{g(-1+\Delta_1-\Delta_2)}\\
    0&0&0&1&0&\dfrac{\pm\epsilon(\Delta_1-\Delta_2)}{g(1+\Delta_1-\Delta_2)}\\
    0&0&0&0&1&\dfrac{\mp2\epsilon(\Delta_1-\Delta_2)}{(-1+\Delta_1-\Delta_2)(\Delta_1+\Delta_1^{2}+\Delta_2-\Delta_2^{2})}\\
    0&0&0&0&0&0\\
    0&0&0&0&0&0\\
    0&0&0&0&0&0\\
  \end{pmatrix}.
\end{small} 
\end{eqnarray}

Therefore, the eigenstate reads
\begin{equation}\label{EE}
\begin{aligned}
|\psi_{R{\epsilon}}\rangle &=\frac{1}{{\cal N}}[\dfrac{\pm(\Delta_1-\Delta_2)(-1+\Delta_1+\Delta_2)}{2}|0,g,g\rangle\\
   &-\dfrac{(\Delta_1-\Delta_2)(1+\Delta_1+\Delta_2)}{2}|0,e,e\rangle\\
   &+\dfrac{\epsilon(\Delta_1-\Delta_2)}{-1+\Delta_1-\Delta_2}|0,e,g\rangle\\
   &\mp\dfrac{\epsilon(\Delta_1-\Delta_2)}{1+\Delta_1-\Delta_2}|0,g,e\rangle\\
   &+g|1\rangle(|e,g\rangle\mp|g,e\rangle)\\
   &\pm\dfrac{2\epsilon(\Delta_1-\Delta_2)}{(-1+\Delta_1-\Delta_2)(\Delta_1+\Delta_1^{2}+\Delta_2-\Delta_2^{2})}\\
   &\times g|1\rangle(|g,g\rangle\mp|e,e\rangle)].
\end{aligned}
\end{equation}
\begin{figure}
\centering
\raisebox{-0.1mm}{\includegraphics[width=0.5\linewidth]{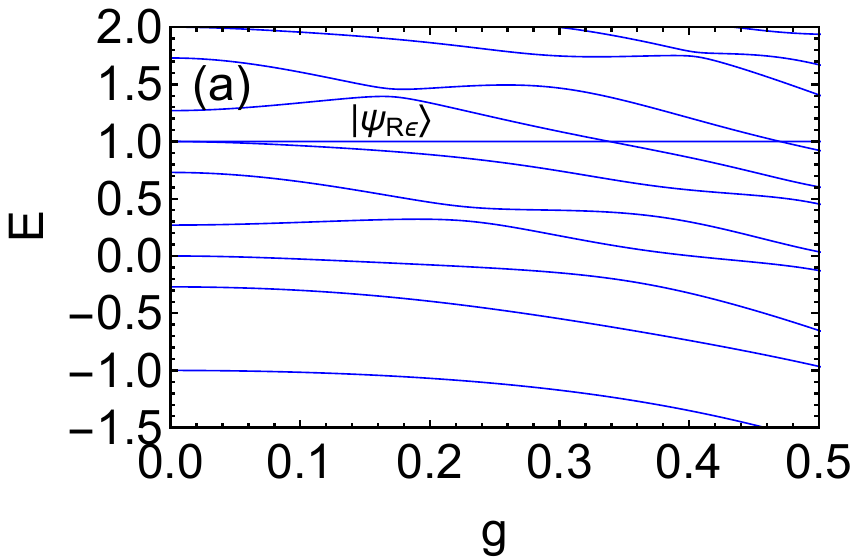}}\hfill
\scalebox{1.0}[0.998]{\includegraphics[width=0.5\linewidth]{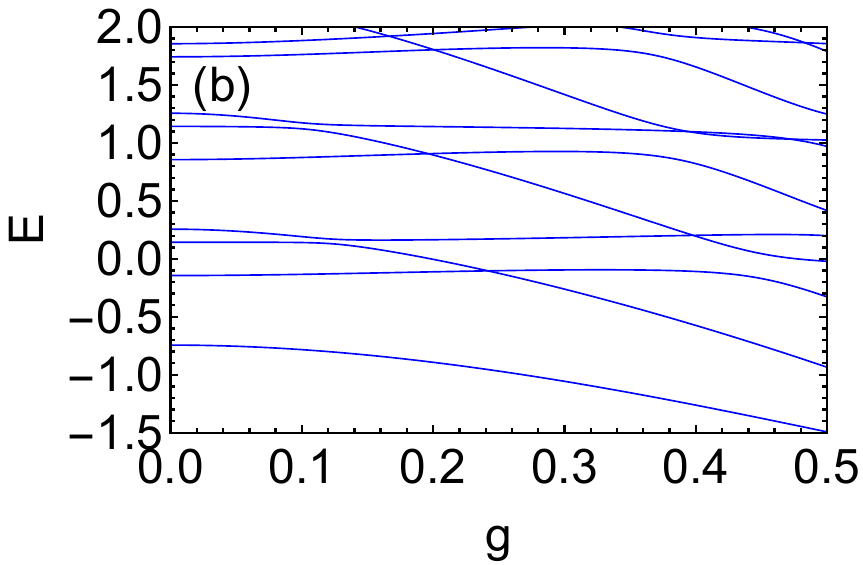}}
\caption{The spectrum of the two-qubit AQRM with (a) $g_1=g_2=g$, $\Delta_1=0.6$, $\Delta_2=0.3$, $\epsilon_1=\epsilon_2=\epsilon=\dfrac{\sqrt{\Delta_1^{4}+(-1+\Delta_2^{2})^{2}-2\Delta_1^{2}(1+\Delta_2^{2})}}{2} = \dfrac{\sqrt{1729}}{200} $, and (b) $g_1=g_2=g$, $\Delta_1=\Delta_2=0.8$, $\epsilon_1=1/2$, $\epsilon_2=0$.}
\label{fig.1}
\end{figure}
Since it has constant energy and exists independent of the relation between $g$ and other parameters, it corresponds to a horizontal line in the spectrum, as shown in fig. \ref{fig.1}(a), while still being a qubit-photon entangled state. Obviously, this horizontal line will bring level crossings, although there is no explicit conserved quantity except energy. Its prominent characteristic is that the level crossings only happens between $|\psi_{R{\epsilon}}\rangle$ and other energy levels, therefore we can label this degeneracy sufficiently with the eigenvalues of $|\psi_{R{\epsilon}}\rangle\langle \psi_{R{\epsilon}}\vert$, $0$ and $1$. This operator obviously commutes with $H$, and has an analytical form. However, we still need an operator contaning operators like $ a $ and $ a^\dagger $, commuting with $H$ to confirm the existence of a hidden symmetry.

According to Ashhab \cite{atf}, symmetric operators can be expressed in the eigenenergy basis as $\hat{S}=\sum_{i,j} s_i \vert \psi_{i,j}\rangle\langle \psi_{i,j}\vert$, where $\psi_{i,j}$ is the $j$-th eigenstate of  $\hat{S}$ with eigenvalue $s_i$. We can obtain the information of $\hat{S}$ from the spectrum. If level crossings only happen between two groups of energy levels, then we only need two $s_i'$s to label the degeneracy. If $s_i=\pm1$, then $\hat{S}$ can be a parity operator, e.g., $\exp{(i\pi a^\dag a)}\sigma_z=\sum_n(\vert \psi_{+,n}\rangle\langle \psi_{+,n}\vert-\vert \psi_{-,n}\rangle\langle \psi_{-,n}\vert)$ for the standard QRM. Or in some cases, $s_i$ is dependent on parameters, and then so does $\hat{S}$, e.g., the hidden symmetry operator $J$ obtained in \cite{ths,GSO} for the AQRM. If level crossings happen between $N$ groups of energy levels, then  $\hat{S}$ should have $N$ eigenvalues. E. g., for the standard Jaynes-Cumming (JC) model, the conserved excitation number operator $C=a^\dag a+(\sigma_z+1)/2=\sum_{i=1,2,3\ldots}i(\vert\psi_{i,+}\rangle\langle\psi_{i,+}\vert+\vert\psi_{i,-}\rangle\langle\psi_{i,-}\vert)$ is obtained by choosing $s_i=0,1,2,3\ldots$. Here level crossings only happen between $|\psi_{R{\epsilon}}\rangle$ and other energy levels, so it is convenient to study its symmetric operator in the eigenenergy basis. We can easily write 
\begin{equation}\label{se}
\hat{S}= \vert \psi_{R{\epsilon}}\rangle\langle \psi_{R{\epsilon}}\vert+f(\Delta_{1,2},\epsilon,g)\sum_{\psi\neq\psi_{R{\epsilon}}}\vert \psi\rangle\langle\psi\vert.
\end{equation} The simplest choice is $f(\Delta_{1,2},\epsilon,g)=0$, where  $\hat{S}$ can have an analytical form, which is still dependent on parameters. However, a symmetric operator written in terms of $a$,  $a^\dag$ or Pauli operators is still needed to confirm the existence of a hidden symmetry. Eq. \eqref{se} may give a hint since it reveals a possible form in the eigenenergy basis.

As discussed in \cite{shi}, there is another kind of level crossings in the two-qubit AQRM, where $\epsilon_1=1/2$, $\epsilon_2=0$,  $g_1=g_2$, $\Delta_1=\Delta_2$, which is depicted in fig. \ref{fig.1}(b). This level crossings is brought about by the hidden symmetry operator \cite{shi}, which reads
\begin{widetext}
\begin{equation}\label{C}
\begin{aligned}
e^{i\pi a^\dagger a}\begin{pmatrix}
   a^\dagger-a+4g+\dfrac{\Delta}{g}&0&a^\dagger+a&0\\
   0&-a^\dagger+a-\dfrac{\Delta}{g}&-4g&-a^\dagger-a\\
   -a^\dagger-a&-4g&-a^\dagger+a+\dfrac{\Delta}{g}&0\\
   0&a^\dagger+a&0&a^\dagger-a+4g-\dfrac{\Delta}{g}\\
  \end{pmatrix}
\end{aligned} 
\end{equation}
\end{widetext}
\begin{floatequation}
\mbox{\textit{see eq.~\eqref{C} below}}
\end{floatequation}
in the qubit basis $\{\vert e,e\rangle,\vert e,g\rangle,\vert g,e\rangle,\vert g,g\rangle\}$.
Different from the former case, here level crossings happen between different energy levels, so it is impossible to label all the degeneracies with a certain $\vert E_n\rangle\langle E_n\vert$.  This operator (eq.~\eqref{C}) is written in the qubit basis and contains $a$ and $a^\dag$.

Actually, hidden symmetries of generalized Rabi models do not only exist in the above case. There are hidden symmetries in the asymmetric N-qubit \cite{shi}, two-mode \cite{ASA}, two-photon \cite{SOO,dda}, anisotropic and the Rabi–Stark model \cite{HSO,hsa}. These level crossings are all brought about by the qubit bias. However, level crossings can also be present even in the absence of the bias. Choosing $\epsilon=0$ in eq. \eqref{EE}, $\vert \psi_{R\epsilon}\rangle$ reduces to
\begin{equation}\label{a}
\begin{aligned}
|\psi_{R}\rangle &=\frac{1}{{\cal N}^{\prime}}[(\Delta_1-\Delta_2)|0,e,e\rangle+g|1\rangle(|g,e\rangle\mp|e,g\rangle)],
\end{aligned}
\end{equation}
with the condition $\Delta_1+\Delta_2=1$, $g_2=\pm g_1=g$, and $ E=1 $, which has been found in \cite{SOT}. This solution still corresponds to a horizontal line in the spectrum and obviously causes level crossings in the parity subspace. So although the parity $\exp{(i\pi a^\dag a)}\sigma_{1z}\sigma_{2z}$ is restored, we still need another conserved quantity (possible hidden symmetry) to label such level crossings. We can construct such operator by $\hat{S}= \vert \psi_{R}\rangle\langle \psi_{R}\vert+f(\Delta_{1,2},g)\sum_{\psi\neq\psi_{R}}\vert \psi\rangle\langle\psi\vert$ in the eigenenergy basis, because level crossings only happen between $\vert\psi_R \rangle$ and other energy levels. However, we have not obtained its analytical form in terms of operators like $a$ and $a^\dag$. 

\begin{figure}
\centering
\scalebox{1.0}[1.0]{\includegraphics[width=0.5\linewidth]{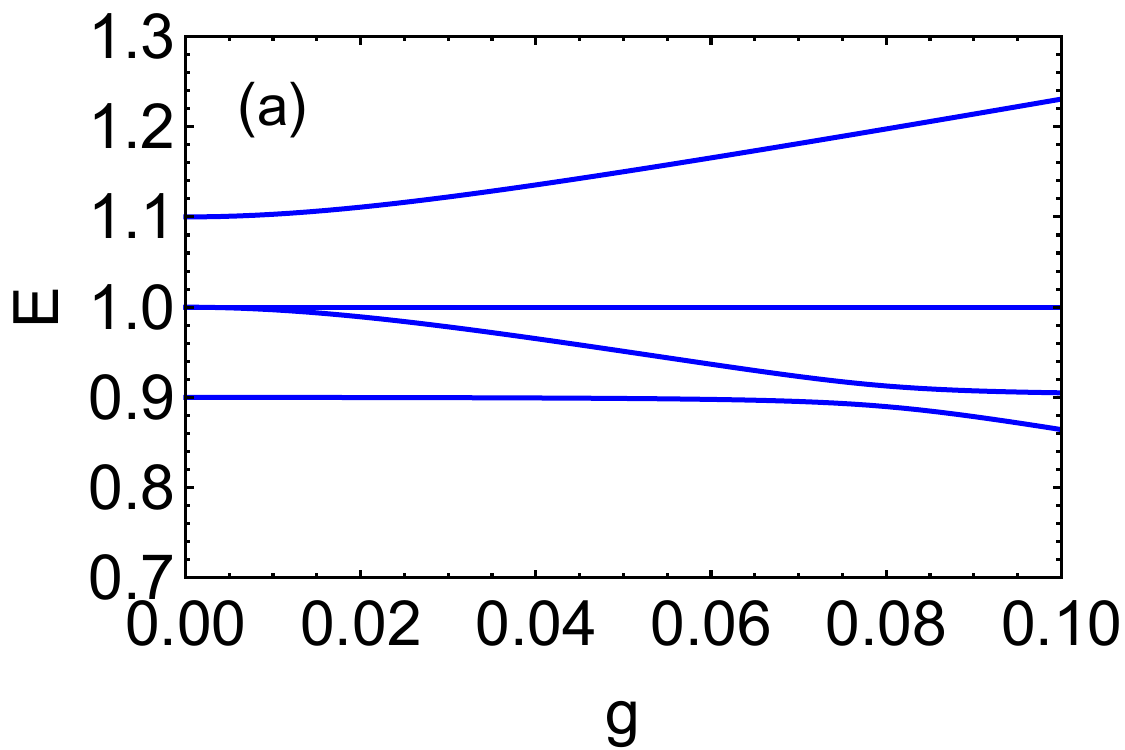}}\hfill
\scalebox{1.0}[1.0]{\includegraphics[width=0.5\linewidth]{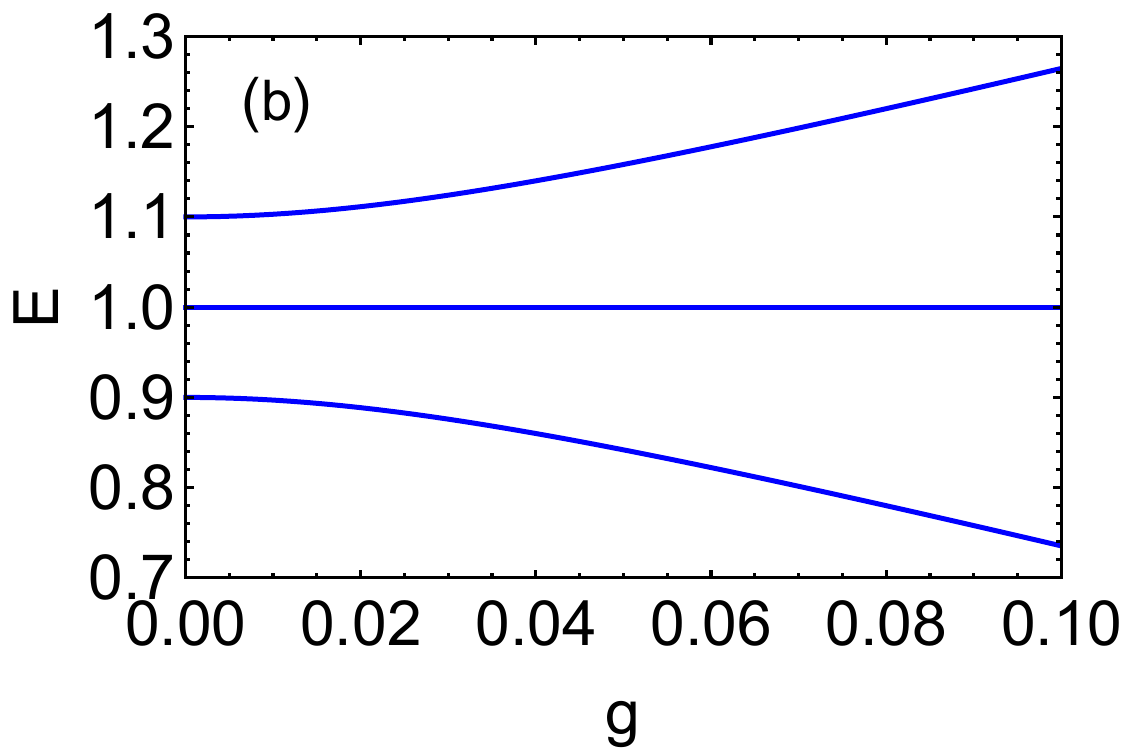}}
\caption{The spectrum of the two-qubit Jaynes-Cummings model in the subspace of $C=2$ with (a) $\Delta_1=0.55$, $\Delta_2=0.45$, $g_2=0.1g_1=0.1g $, and (b) $\Delta_1=0.55$, $\Delta_2=0.45$, $g_2=g_1=g$.}
\label{fig.2}
\end{figure}

Interestingly, such degeneracies also happen in the subspace of the two-qubit JC model
\begin{equation}\label{2jc}
H=\omega a^\dag a+\sum_{i=1,2}(g_i a\sigma_{i}^\dag+g_i a^\dag\sigma_i+\Delta_{i}\sigma_{iz}).
\end{equation}
The excitation number operator $C=a^\dag a +(\sum_{i=1,2}\sigma_{iz}+2)/2$ is conserved. The detunings of the photon frequency from the qubit transition frequencies are defined as $\delta_{1,2}=2\Delta_{1,2}-\omega$.  When $\delta_1+\delta_2=0$, i.e., $\Delta_1+\Delta_2=\omega$, there are dark states
\begin{equation}
\vert \psi_d\rangle=\sqrt{N+2}\vert N,e,e\rangle-\sqrt{N+1}\vert N+2,g,g\rangle
\end{equation}
with constant eigenenergies $E=N+1$ ($N=0,1,2,\ldots$) in the whole coupling regime, as shown in fig. \ref{fig.2} (a) for $N=0$. This is because the superposition of $\vert N,e,e\rangle$ and $\vert N+2,g,g\rangle$ could cancel the population of $\vert N+1,e,g\rangle$ and $\vert N+1,g,e\rangle$ when applied by $H$ (Eq. \eqref{2jc}). Although $\vert \psi_d\rangle$ has constant eigenenergy, it will not cause level crossings. On the other hand,  $H(\vert N+1,e,g\rangle\mp\vert N+1,g,e\rangle)$ only involves basis  $\vert N+1,e,g\rangle$ and $\vert N+1,g,e\rangle$ when $g_1=\pm g_2$. If $\Delta_1=\Delta_2$, then $\vert N-1,e,g\rangle\mp\vert N-1,g,e\rangle$ is a dark-state solution with the same constant energy  as
$\vert \psi_d\rangle$, which causes degeneracies in the subspace of $C=N+2$. It can be explained from the conserved total angular momenta for the spin singlet state $\vert N+1,e,g\rangle-\vert N+1,g,e\rangle$. Meanwhile, if $\Delta_1+\Delta_2=\omega$, then 
\begin{equation}
\vert \psi_{d1}\rangle=(\Delta_1-\Delta_2)\vert N,e,e\rangle+\sqrt{N+1}g\vert N+1\rangle(\vert g,e\rangle\mp\vert e,g\rangle)
\end{equation}
becomes a dark state solution to $H$ (eq. \eqref{2jc}) with the same constant energy as $\vert \psi_d\rangle$. They degenerate with each other in the subspace of $C=N+2$, as shown in fig. \ref{fig.2} (b) for $N=0$. If the counter-rotating terms are further included, $\vert \psi_{d1}\rangle$ is still a dark state solution for $N=0$, which reduces to $\vert \psi_R\rangle$. To conclude, the possible hidden symmetry is close related to the resonance condition $\delta_1+\delta_2=0$ and partially broken permutation
 symmetry ($g_1=\pm g_2$, $\Delta_1\neq\Delta_2$). It is interesting to note that flat band may arise due to symmetries or fine tuning of lattice parameters \cite{Daniel,dirac}, just like the dark state here, so they may have a link.

\section{Extended to the multimode case}\label{s3}
The multimode two-qubit AQRM reads
\begin{eqnarray}\label{h2}
H_M&=&\sum_{i=1}^{M}\omega_ia_i^\dagger a_i+\Delta_1\sigma_{1z}+\Delta_{2}\sigma_{2z}\nonumber\\
&&+\sum_{i=1}^{M}(g_{i1}\sigma_{1x}+g_{i2}\sigma_{2x})(a_i^\dagger+a_i)\nonumber\\
&&+\epsilon_{1}\sigma_{1x}+\epsilon_{2}\sigma_{2x},
\end{eqnarray} 
where $a_i^\dagger$ and $a_i$ are the $ i $-th photon mode creation and annihilation operators with frequency $\omega_i$, respectively. $ g_{i1}$ and $g_{i2}$ are the qubit-photon coupling strength between the $ i $-th mode and two qubits, respectively.

When $ \omega_i= \omega $ and $  g_{i1}/g_{i^\prime 1}=g_{i2}/g_{i^\prime 2}$, we can introduce similar Bogoliubov operators as proposed in \cite{stt}
\begin{equation}
\label{E5}
\begin{split}
b_1&=\dfrac{\sum_{i=1}^{M}g_{i1}a_i}{\sqrt{\sum_{i=1}^{M}g_{i1}^{2}}},\\
b_j&=\dfrac{\sum_{i=1}^{j-1}g_{i1}g_{j1}a_i-\sum_{i=1}^{j-1}g_{i1}^{2}a_j}{\sqrt{\sum_{i=1}^{j}g_{i1}^{2}\sum_{i=1}^{j-1}g_{i1}^{2}}}, j=2,3,\dots,M
\end{split}
\end{equation}
to transform eq. \eqref{h2} into
\begin{eqnarray}\label{E71}
 H_M^{\prime}&=&\omega b_1^\dagger b_1+\Delta_{1}\sigma_{1z}+\Delta_{2}\sigma_{2z}\nonumber\\
 &&+\left[(\sum_{i=1}^{M}g_{i1}^{ 2})^\frac{1}{2}\sigma_{1x}+(\sum_{i=1}^{M}g_{i2}^{ 2})^\frac{1}{2}\sigma_{2x}\right](b_1^\dagger+b_1)\nonumber\\ 
 &&+\epsilon_{1}\sigma_{1x}+\epsilon_{2}\sigma_{2x}+\omega\sum_{j=2}^{M}b_j^\dagger b_j.
\end{eqnarray}

$H_M^{\prime}$ is a combination of the two-qubit AQRM and $M-1$ free bosonic modes. So its solution takes the form of $\vert \psi_{b_1}\prod_{j=2}^{M} n_{b_j}\rangle$. $\vert n_{b_j}\rangle$ is the eigenstate of $b_j^\dag b_j$ and $\vert\psi_{b_1}\rangle$ can be obtained from the solution of the single mode case by replacing $g_1$ with $(\sum_{i=1}^M g_{i1}^{ 2})^\frac{1}{2}$, $g_2$ with $(\sum_{i=1}^M g_{i2}^{ 2})^\frac{1}{2}$ and $a$ by $b_1$. Dark-state solution to the two-qubit AQRM $|\psi_{R{\epsilon}}\rangle$ exists when $g_1=g_2=g$, so a similar solution $|\psi_{R{\epsilon}}^{\prime}\rangle$ for the multimode case also requires $(\sum_{i=1}^M g_{i1}^{ 2})^\frac{1}{2}=(\sum_{i=1}^M g_{i2}^{2})^\frac{1}{2}=g_b$. Considering $  g_{i1}/g_{i^\prime 1}=g_{i2}/g_{i^\prime 2}$, we arrive at $g_{i1}=g_{i2}=g^\prime_{i}$. $|\psi_{R{\epsilon}}^{\prime}\rangle$ can be obtained by replacing $\vert 1\rangle=a^\dag\vert 0\rangle$ with $ |W_{M}\rangle =b_1^\dag\vert 0\rangle= (\sum_{i=1}^{M}{g^\prime_{i}}^{2})^{-\frac{1}{2}}\sum_{i=1}^{M}g_{i}^{\prime}|0_{1}0_{2}\dots1_{i}0_{i+1}\dots0_{M}\rangle$,  and $g$ with $g_b$ in $|\psi_{R{\epsilon}}\rangle$ (eq. \eqref{EE}).  $|\psi_{R{\epsilon}}^{\prime}\rangle$ reads

\begin{figure}
\begin{center}
\raisebox{0.5mm}{\includegraphics[width=0.5\linewidth]{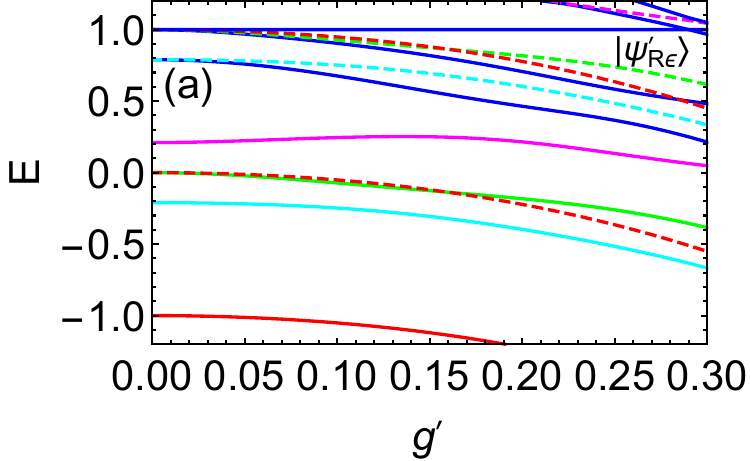}}\hfill
\scalebox{1.0}[0.996]{\includegraphics[width=0.5\linewidth]{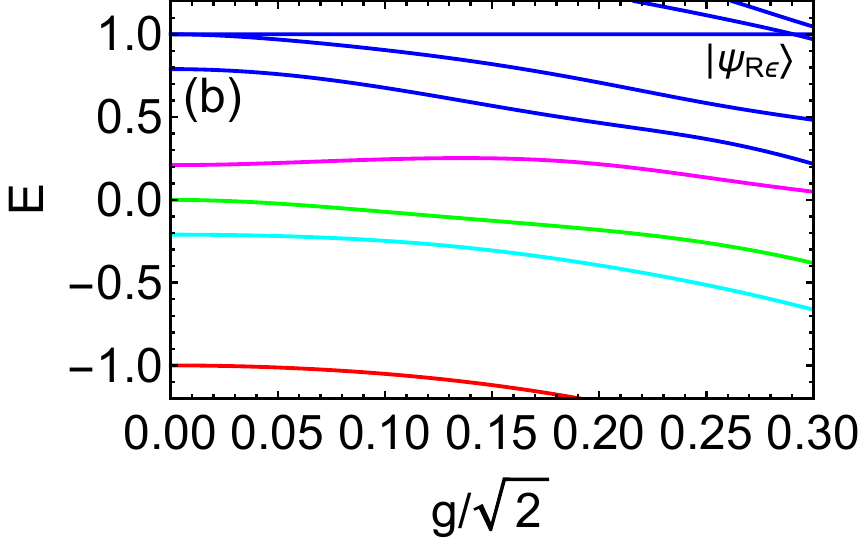}}
\raisebox{-0.01mm}{\includegraphics[width=0.5\linewidth]{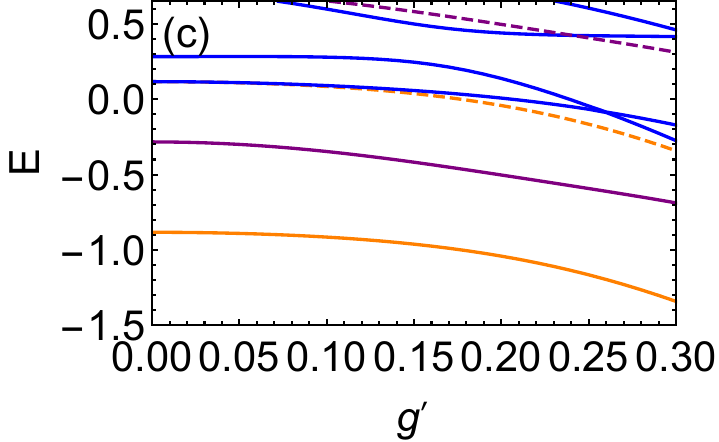}}\hfill
\scalebox{1.0}[0.98]{\includegraphics[width=0.5\linewidth]{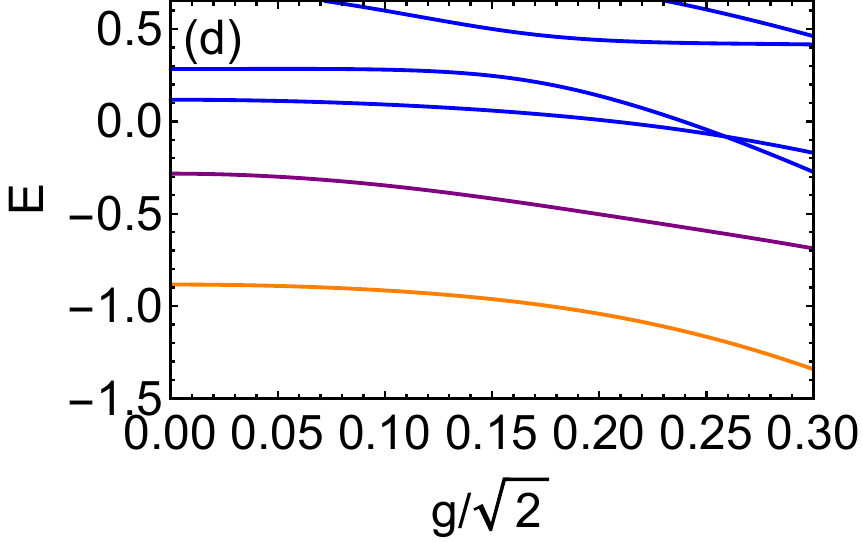}}
\caption{(a) and (c) Spectrum of the two-qubit two-mode AQRM with $\omega_1=\omega_2=\omega=1$, $g_1^\prime=g_2^\prime=g^\prime$. The former has $\Delta_1=0.5$, $\Delta_2=0.2$, $\epsilon_1=\epsilon_2=\epsilon=\dfrac{\sqrt{4641}}{200}$. The latter has $\Delta_1=\Delta_2=0.3$, $\epsilon_1=1/2$, $\epsilon_2=0$. (b) and (d) Spectrum of the two-qubit single-mode AQRM with parameters the same as in (a) and (c) respectively, except $g=\sqrt{2}g^{\prime}$.}
\label{fig.3}
\end{center}
\end{figure}

\begin{equation}\label{EEE}
\begin{aligned}
|\psi'_{R{\epsilon}}\rangle &=\frac{1}{{\cal N^{''}}}[\dfrac{\pm(\Delta_1-\Delta_2)(-1+\Delta_1+\Delta_2)}{2}|0_M,g,g\rangle\\
   &-\dfrac{(\Delta_1-\Delta_2)(1+\Delta_1+\Delta_2)}{2}|0_M,e,e\rangle\\
   &+\dfrac{\epsilon(\Delta_1-\Delta_2)}{-1+\Delta_1-\Delta_2}|0_M,e,g\rangle\\
   &\mp\dfrac{\epsilon(\Delta_1-\Delta_2)}{1+\Delta_1-\Delta_2}|0_M,g,e\rangle\\
   &+g_b|W_M\rangle(|e,g\rangle\mp|g,e\rangle)\\
   &\pm\dfrac{2\epsilon(\Delta_1-\Delta_2)}{(-1+\Delta_1-\Delta_2)(\Delta_1+\Delta_1^{2}+\Delta_2-\Delta_2^{2})}\\
   &\times g_b|W_M\rangle(|g,g\rangle\mp|e,e\rangle)].
\end{aligned}
\end{equation}

It still has constant eigenenergy $E=1$, corresponding to a horizontal line in the spectrum, as shown in fig. \ref{fig.2} (a) for $M=2$. Such a horizontal line will obviously cause level crossings. Its degeneracy can be labelled by the eigenvalues of  $|\psi_{R{\epsilon}}^{\prime}\rangle\langle\psi_{R{\epsilon}}^{\prime}\vert$, $0$ and $1$. Meanwhile, the dashed lines with $n_b=\sum_{j}n_{b_j}\neq0$ are translations of the same-color solid lines with $n_b=0$ by $n_b\omega$, according to eq. \eqref{E71}.  This will cause another kind of level crossings between energy levels with different $n_b$, as found in the multimode QRM without bias \cite{stt}. The corresponding symmetry operator reads $b^\dag_j b_j$, and can be used to label the degeneracies by its eigenvalues $n_{b_j}=0,1,2,\ldots$.

On the other hand, the spectrum of the two-qubit multimode AQRM is the same as the single mode two-qubit AQRM except for the energy levels with nonzero $n_{b_j}$, if we choose $g_b=g$, according to eq. \eqref{E71}. If $g_1^\prime=g_2^\prime=g^\prime$ for the two-mode case, then the corresponding $g=\sqrt{2}g^\prime$ for the single mode case. Then the spectra will be the same for $n_b=0$, as shown in figs. \ref{fig.3} (a) and (b). If we choose $\Delta_1=\Delta_2$, $\epsilon_1=1/2$, $\epsilon_2=0$, the level crossings in the single mode AQRM also appears in the multimode case, as shown in figs. \ref{fig.3} (c) and (d). The hidden symmetry operator can be obtained simply by replacing $a$ with $b_1$ and $g$ with $g_b$ in eq. \eqref{C}. There is also another kind of level crossings between dashed lines and other energy levels, caused by the symmetry operator  $b^\dag_j b_j$, since it commutes with the Hamiltonian eq. \eqref{E71}.

\section{ Conclusion}\label{s4}
We study the two-qubit AQRM and find its dark-state solution with at most one photon and constant energy in the whole coupling regime, which causes level crossings, although there is no explicit conserved quantity except energy. We find an operator in the eigenenergy basis to label all the degeneracies and compare it with the hidden symmetry operator found in \cite{shi}. The level crossings brought about by the dark-state solution still exists in the parity subspace when there is no qubit bias. Whether such level crossings due to a hidden symmetry or just fine tuning of parameters like the contacts of energy bands \cite{Daniel,dirac} is an interesting question to explorer. One evidence supporting the former is that the adiabatic evolution along the dark state is successful in the vicinity of the crossing point, as if ``protected'' by certain symmetry, just like the case of $\epsilon=n\omega/2$, where the adiabatic evolution along a certain energy level is also successful although there are level crossings during the process.  Extended to the two-qubit $M$-mode AQRM, there are $M-1$ conserved bosonic number operator $b_j^\dag b_j$ for $j=2,3,\ldots,M$.  Such symmetries also cause level crossings in the $M$-mode AQRM. This work  develops a perspective on the symmetry studies of generalized Rabi models.

\acknowledgments
This work was supported by the Scientific Research Fund of Hunan Provincial Education Department (Grants No. 23A0135), Natural Science Foundation of Hunan Province, China (Grants No. 2022JJ30556, No. 2023JJ30596, and No. 2023JJ30588), National Natural Science Foundation of China (Grants No. 11704320).

\end{document}